\documentclass[%
amsmath,amssymb,
 reprint,%
aip,
cha
]{revtex4-1}

\usepackage{amssymb}
\usepackage{graphicx}
\usepackage{dcolumn}
\usepackage{bm}

\usepackage[english]{babel}
\usepackage{lipsum}
\usepackage[]{longtable}

\usepackage[dvipsnames]{xcolor}

\begin{document}


\title{From neurons to epidemics: How trophic coherence affects spreading processes}

\author{Janis Klaise}

\affiliation{Centre for Complexity Science,
University of Warwick,
Coventry CV4 7AL, United Kingdom.}

\author{Samuel Johnson}
\email[]{s.johnson.2@warwick.ac.uk}

\affiliation{Centre for Complexity Science,
University of Warwick,
Coventry CV4 7AL, United Kingdom.}
\affiliation{Warwick Mathematics Institute,
University of Warwick,
Coventry CV4 7AL, United Kingdom.}

\begin{abstract}
Trophic coherence, a measure of the extent to which the nodes of a directed network are organised in levels,
has recently been shown to be
closely related to many structural and dynamical aspects of complex systems, including graph eigenspectra,
the prevalence or absence of feedback cycles, and linear stability.
Furthermore, non-trivial trophic structures have been observed in networks of neurons, species, genes, metabolites,
cellular signalling, concatenated words, P2P users, and world trade.
Here we consider two simple yet apparently quite different dynamical models -- one a Susceptible-Infected-Susceptible 
(SIS) epidemic model adapted to include complex contagion, the other an Amari-Hopfield neural network --
and show that in both cases the related spreading processes are modulated in similar ways by the trophic coherence
of the underlying networks. To do this, we propose a network assembly model which can generate structures with 
tunable trophic coherence, limiting in either perfectly stratified networks or random graphs.
We find that trophic coherence can exert a qualitative change in spreading behaviour, determining 
whether a pulse of activity will percolate through the entire network or remain confined to a
subset of nodes, and whether such activity will quickly die out or endure indefinitely.
These results could be important for our understanding of phenomena such as epidemics, rumours, shocks to ecosystems, 
neuronal avalanches, and many other spreading processes.
\end{abstract}

\pacs{89.75.Fb, 87.18.-h, 64.60.ah, 84.35.+i, 87.23.-n}






\maketitle 

\begin{quotation}
A great many processes involve some kind of activity travelling through a complex system:
a rumour or contagious disease in society, waves of action potentials between neurons, 
cascades of defaults in a banking system, or species under stress in an ecosystem, for instance.
It is well-known that food webs -- or networks of predation -- have a trophic structure, a hierarchy 
of plants, herbivores, omnivores, primary carnivores, and so on up to top predators, defined by how many steps separate
each species from the source of energy. It has recently been shown that ``trophic coherence'', a measure of
how neatly these nodes fall into distinct layers, is key to understanding many structural and dynamical 
features not just of food webs, but of a range of biological and artificial networks.
In light of these observations, we now study how trophic coherence affects spreading processes,
and find that it plays an important role. Our results suggest, for example, that whether a contagious disease 
will become endemic, or whether a sensory stimulus will propagate through a  brain,
could be determined by the trophic coherence of the underlying networks.
\end{quotation}

\section{Introduction}

 \begin{figure*}
 \includegraphics[scale=1]{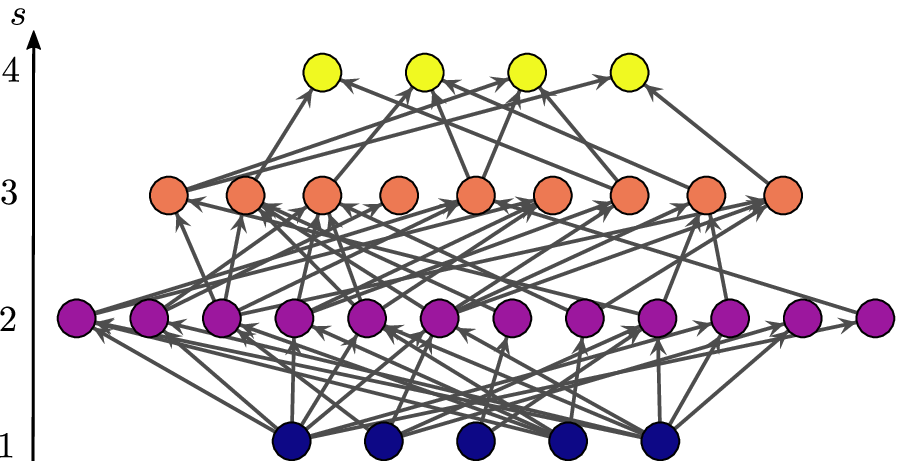}
  \includegraphics[scale=1]{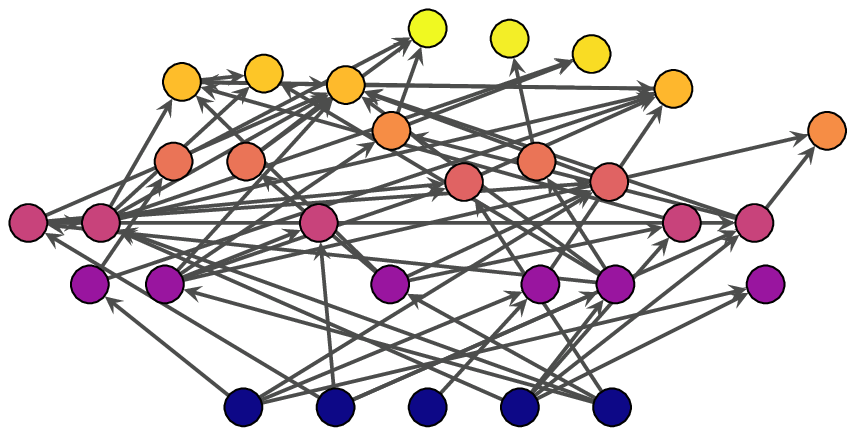}
 \caption{Left: An example of a maximally coherent network ($q=0$). Right: A network with the same parameters 
 $N$, $B$ and $L$ as the one on the left, but less trophically coherent ($q=0.49$). 
 In both cases, the hight of the nodes on the vertical axis represents their trophic level.
 The networks were generated with the preferential preying model as described in the main text, 
 with $T=0.001$ for the one on the 
 left, and $T=1$ for the one on the right.
 }%
 \label{fig_pic1}
 \end{figure*}

\noindent
An activist tweets a picture and the image spreads through social networks, perhaps going viral;
information enters the brain via sensory neurons and cascades through various kinds of cell before 
inducing motor neurons to fire; 
the plants in an ecosystem are decimated by a 
drought
and the pulse of want 
travels up through trophic chains until even the 
top predators are affected. These and many others are examples of spreading processes in complex systems which can be regarded as
cascades of activity percolating through directed networks. Crucially, the signals arriving at a node -- person, neuron, species -- 
can combine to create a nonlinear response: it is not the same to receive a trickle of signals over an extended period of time 
as a single, large package.

Over the past decade and a half much work has gone into studying the effect of network topology 
on dynamical processes of various kinds.\cite{boccaletti2006complex,arenas2008synchronization,barrat2008dynamical}
One of the first widely acknowledged consequences of the small-world property 
famously described by Watts and Strogatz\cite{WattsStrogatz}
was that epidemics could spread rapidly through social networks because of just a few 
long-range connections.\cite{moore2000epidemics} 
Subsequent work on epidemic models
has revealed a rich relationship between network 
structure and 
spreading.\cite{keeling_networks_2005,durrett_features_2010,danon_networks_2011,house_modelling_2012,pastor-satorras_epidemic_2015}
For instance, the phenomenon of `complex contagion', which Centola showed experimentally to play a part in online social networks,\cite{Centola1194}
has recently been investigated mathematically on clustered network models.\cite{osullivan_mathematical_2015}
Network structure is also important in models of opinion formation,\cite{suchecki2005voter,sood2008voter}
and in this context the `q-voter model' captures the idea of complex contagion.\cite{castellano2009nonlinear,moretti2013mean}
Activity on neural networks, while not usually studied as spreading processes, has also been shown, in simple models,
to depend fundamentally 
on topological properties, such as degree heterogeneity,\cite{johnson2008functional} assortativity,\cite{de2011enhancing} 
and clustering.\cite{johnson2013robust}
Furthermore, much research has focused on mapping the structure of biological neural networks, and understanding how 
such topologies arise.\cite{white1986structure,Song2005,honey2007network,johnson2010evolving,perin2011synaptic,Perin2013}

`Trophic coherence', a measure of how neatly food webs or other directed networks fall into well-defined trophic levels, 
has recently been shown to play a key role in the dynamical stability of ecosystems.\cite{Johnson_trophic}
Such is the influence of this topological property that sufficiently coherent networks can become 
more stable with increasing size and complexity, thereby offering a solution to May's paradox.\cite{May,May_book,Debate,Johnson_trophic}
It has also been suggested that the trophic coherence of ecosystems may have contributed to the devastating effect that human expansion 
had on Pleistocene megafauna.\cite{Pires20151367}
But it is not only food webs which exhibit a significantly non-random trophic structure: this characteristic has recently been shown
to determine properties such as eigenspectra,\cite{Johnson_spectra} feedback,\cite{Vir_loops,Johnson_spectra} directionality,\cite{Vir_loops}
and intervality\cite{Vir_Chaos} in a wide variety of biological and artificial networks, including those of gene transcription,
neurons, metabolites, cellular signalling, concatenated words, P2P users and world trade.\cite{Vir_loops,Johnson_spectra}

In this paper we show that trophic coherence also affects spreading processes. We consider two seemingly rather different models, 
one a version
of the Susceptible-Infected-Susceptible (SIS) epidemic model\cite{pastor-satorras_epidemic_2015} extended to account for complex 
contagion,\cite{castellano2009nonlinear,Centola1194} and the other an Amari-Hopfield 
neural network.\cite{amari1972characteristics,hopfield1982neural,amit1992modeling}

To generate directed networks with tunable trophic coherence we propose a variation of the `preferential preying' model (PPM)
used by Johnson {\it et al.}\cite{Johnson_trophic}, with the difference that this version limits in random graphs instead of acyclic 
`cascade model' networks, and is therefore less specific to food webs and perhaps more appropriate for investigating generic directed networks.
Our numerical study reveals that trophic coherence can determine whether a pulse of activity in either model will propagate through the 
entire network or remain confined to a small fraction of nodes, or become endemic as opposed to dying out soon after initiation. We conclude with some open 
questions and our assessment of the main areas where theoretical research might provide important insights.

\section{Methods}

\subsection{Trophic coherence}
\noindent
Let us consider a directed network given by the $N\times N$ adjacency matrix $A$, with elements $a_{ij}=1$ if there 
is a directed edge from node $j$ to node $i$, and $a_{ij}=0$ if not.
The in- and out-degrees of node $i$ are $k_{i}^{in}=\sum_j a_{ij}$ and $k_{i}^{out}=\sum_j a_{ji}$, respectively.
The number of edges is $L=\sum_{ij}a_{ij}$, and the mean degree is $\langle k\rangle=L/N$.
We shall assume that 
the network is weakly connected, and that
there is a number $B>0$ of nodes with $k^{in}=0$, which we shall refer to as {\it basal nodes}.
It is standard in ecology to define the {\it trophic level} $s_i$ of nodes as
\begin{equation}
s_{i} = 1+\frac{1}{k_{i}^{in}}\sum_j a_{ij}s_{j},
\label{eq_si}
\end{equation}
if $k_{i}^{in}>0$, or $s_i=1$ if $k_{i}^{in}=0$.
In other words, the trophic level of basal nodes (autotrophs in the ecological context) is $s=1$ by convention, 
while other nodes (consumer species) are assigned the mean trophic level of their in-neighbours (resources), plus one. 
\cite{Levine_levels}
Thus, for any directed network in which every node is on at least one directed path originating at a basal node, the 
trophic level of each node is a topological feature easily obtained by solving the linear system of Eq. (\ref{eq_si}).
In a recent paper, Johnson {\it et al.}\cite{Johnson_trophic} characterise each edge in an network with a 
{\it trophic distance}:
$x_{ij}=s_i-s_j$
(not a distance in the mathematical sense since it can take negative values).
They then consider the distribution of trophic distances 
over the network, $p(x)$, which will always have mean
$\langle x\rangle=1$.
The homogeneity of $p(x)$ is called {\it trophic coherence}: the more similar the trophic distances 
of all the edges, the more coherent the network.
As a measure of coherence, one can simply use the 
standard deviation of $p(x)$, which is referred to as an {\it incoherence parameter}: $q=\sqrt{\langle x^2\rangle-1}$.

When applied to food webs -- networks of who eats whom in an ecosystem -- it turns out that trophic coherence is the 
best statistical predictor of linear stability.\cite{Johnson_trophic} 
Furthermore, a simple model which generates networks with 
tunable coherence shows that this property can allow for systems to become more stable with size and edge density --
which suggests a solution to May's paradox, or why large, complex ecosystems are observed to be the most 
stable.\cite{May,May_book,Debate,Johnson_trophic}
By defining the {\it coherence ensemble}, it has recently been shown\cite{Johnson_spectra} that 
trophic coherence is a key factor determining the cycle structure and distribution of eigenvalues in directed 
networks, and that it is possible to compare the empirical value of $q$ for a given network to its random 
expectation,
$\tilde{q}=\sqrt{L/L_B-1}$, where 
$L_B$ is the number of edges connected to basal nodes.
It turns out that certain kinds of biological networks, such as food webs, some gene regulatory networks, or
the {\it C. elegans} neural network, are significantly coherent ($q<\tilde{q}$); while others, most notably metabolic networks, 
are less coherent than the random expectation ($q>\tilde{q}$).\cite{Johnson_spectra}

Figure \ref{fig_pic1} displays two small networks which differ only in their trophic coherence: 
the one on the left is maximally 
coherent ($q=0$), while the one on the right is more incoherent ($q=0.49$).
Even the one on the right, however, is more coherent than the corresponding random expectation ($\tilde{q}=2.24$).
Note that the difference between the two networks is apparent thanks to the nodes being plotted on a vertical axis representing
trophic level -- when this information is not highlighted in the visualisation, significantly coherent networks can appear 
no different from incoherent ones to the naked eye.
From the network on the left, it is clear that a maximally coherent network is also multipartite (or bipartite in the case 
of having only two trophic levels).
Thus, trophic coherence can be regarded as the extent to which a network approaches this state of order, much as modularity 
is a measure of how close an undirected network is to being disconnex.\cite{Newman_rev}

It is not yet known what mechanisms lead to trophic coherence -- or incoherence -- in directed networks. One possibility is that
edges are formed preferentially between nodes with specific functions, which in turn are correlated with trophic level. 
For instance, in an ecosystem the biomass is produced by the plants and flows up the food chain through herbivores, primary carnivores, 
and secondary carnivores until it reaches the apex predators. Likewise, information enters a neural network via sensory neurons, 
and is passed on to interneurons and other cells with various processing functions before reaching the motor neurons.
On the other hand, since trophic coherence is related to several other structural and dynamical properties, it may be that 
in certain systems topologies are selected for their stability, robustness or feedback characteristics, and coherence 
is a secondary effect. Be this as it may, the networks observed to exhibit this feature tend to 
be involved in the transport of some quantity -- such as energy or information -- through a system.
In this paper we therefore explore the effects of trophic coherence on spreading processes, and find that it plays an 
important role, in some circumstances inducing transitions between qualitatively different regimes of dynamical behaviour.

 \begin{figure}
 \includegraphics[scale=1]{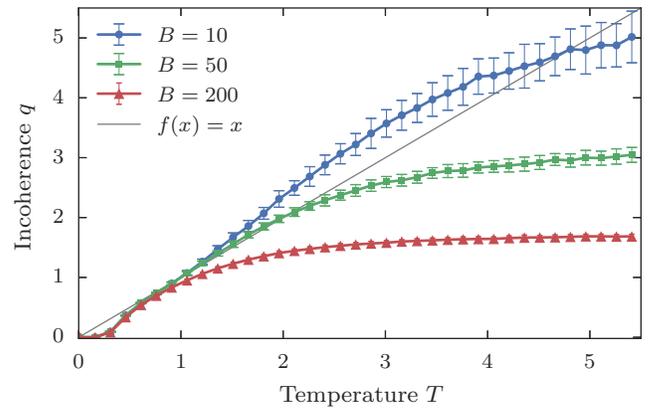}%
 \caption{Trophic coherence, as given by $q$, against the temperature parameter $T$ for networks generated with the 
preferential preying model described in the main text, for different numbers of basal nodes: $B=10$, $50$ and $200$, as shown.
In all cases, the number of nodes is $N=1000$ and the mean degree is $\langle k\rangle = 5$.
Averages are over $1000$ runs.}
 \label{fig_qvt}
 \end{figure}

\subsubsection{Generating coherent networks}
\noindent
Johnson {\it et al.}\cite{Johnson_trophic} put forward a model for generating networks with a tunable degree of trophic 
coherence, referred to as the `preferential preying model' (PPM). 
However, the networks thus generated are always acyclic.
While this is a characteristic of many food webs, we are here interested in 
studying the effects of trophic coherence on spreading processes in the most general circumstances possible, and so
propose an extension of the PPM which can generate cycles. In the first step of the model, we begin with $B$ basal 
nodes and proceed to introduce 
$N-B$ non-basal nodes sequentially; each of these has, at this stage, only one in-neighbour, chosen randomly 
from among the extant nodes (basal and non-basal) already in the network when it arrives. 
At the end of this step, each node $i$ has a preliminary trophic level, $\tilde{s}_i$, as given by Eq. (\ref{eq_si})
(in every case an integer, since each non-basal node has so far only one in-neighbour).
The second step is to introduce
the remaining $L-N+B$ edges needed to make up a total of $L$. For this, each pair of nodes $i,j$ such that $i$ is a non-basal node
is attributed a tentative trophic distance $\tilde{x}_{ij}=\tilde{s}_i-\tilde{s}_j$. Edges between pairs are then placed with 
a probability proportional to
\begin{equation}
P(a_{ij}\rightarrow 1)\propto \exp\left[-\frac{(\tilde{x}_{ij}-1)^2}{2T^2}\right], 
\label{eq_Plink}
\end{equation}
until there are $L$ edges in the network, 
and therefore a mean degree of $\langle k\rangle = L/N$.
As in the original PPM, the `temperature' parameter $T$ tunes the degree of
trophic coherence,
with $T=0$ yielding maximally coherent networks ($q=0$), and incoherence increasing monotonically with $T$.
The specific choice for the edge probability is arbitrary, but the form in Eq. (\ref{eq_Plink}) is conducive to 
a Gaussian distribution of distances $x$, which we have found to be a good fit to empirical data on several kinds of 
networks. The relationship between $T$ and $q$ in networks thus generated is shown in Fig. \ref{fig_qvt}. At low 
$T$ we observe that $q\simeq T$, while the coherence saturates to the randomly expected value at higher $T$.

The main difference between the model described above and the original PPM as implemented by 
Johnson {\it et al.}\cite{Johnson_trophic} -- 
and also by Dom{\'i}nguez-Garc{\'i}a {\it et al.} in this same issue\cite{Vir_Chaos} -- is in the networks
generated at high $T$. Both models generate maximally coherent networks when $T\rightarrow 0$. However, 
when $T\rightarrow \infty$, the original version coincides with both the `generalised cascade model', and with 
the `generalised niche model' when its parameter $c=0$, a limiting behaviour 
which makes for useful comparisons in the context of food webs.\cite{Cohen_book,Stouffer_GNM} 
(These food-web models are based on a `niche axis', defined by a random number given at the outset to each node:
in the generalised cascade model, the in-neighbours of node $i$ are attributed randomly from nodes with lower
niche values than $i$; the `niche model' imposes the additional constraint that in-neighbours must be contiguous on the 
niche axis; and the generalised niche model interpolates between the two with a parameter $c$.)
The version of the PPM which we now put forward generates networks which approach directed 
Erd\"os-R\'enyi random 
graphs at high temperature. In fact, at $T\rightarrow \infty$, the networks generated are the overlap of
two graphs: a ``skeleton'' with $N-B$ edges, which ensures that the network is weakly connected; and a 
directed random graph with $L-N+B$ edges, and the constraint that $B$ nodes have $k^{in}=0$.

 \begin{figure*}
 \includegraphics[scale=1]{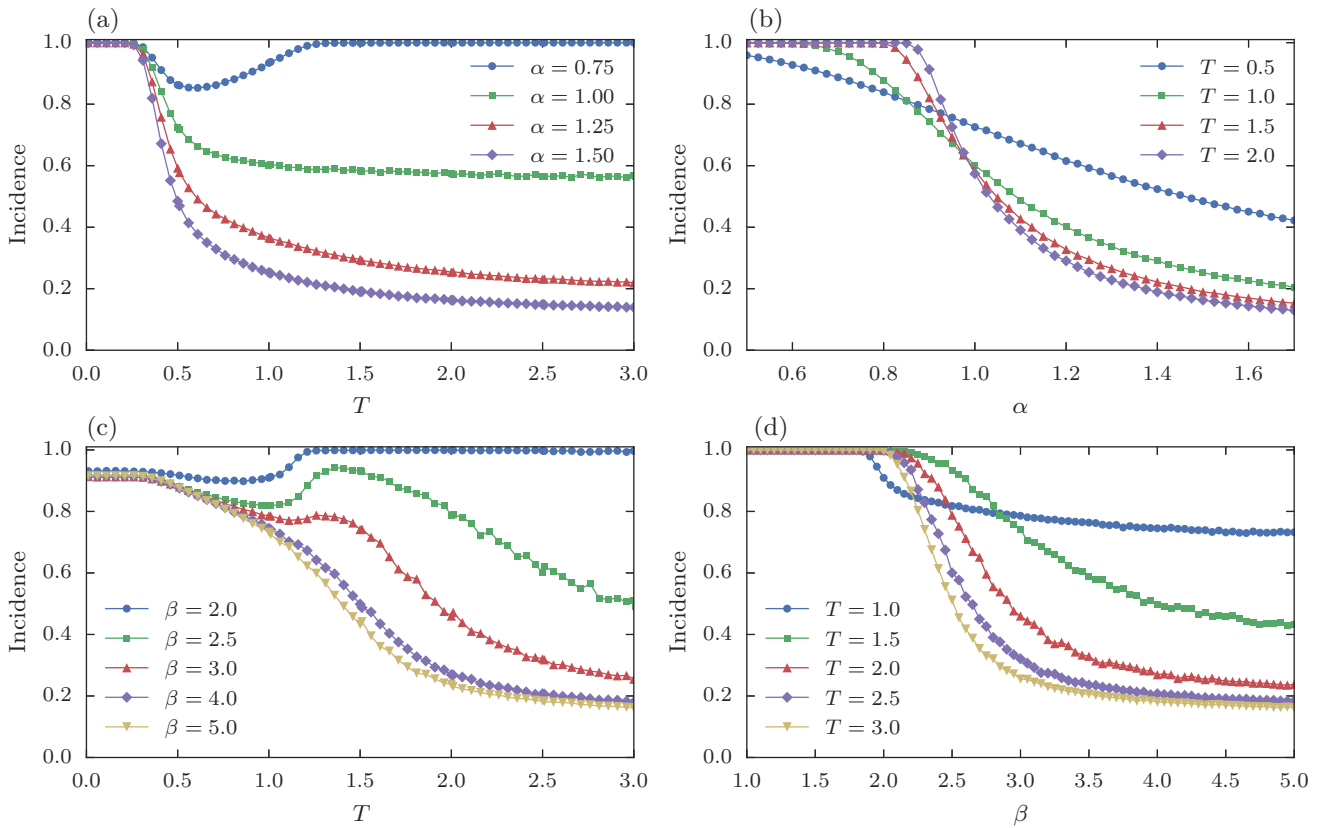}%
 \caption{Average Incidence values from Monte Carlo simulations of the two spreading models on networks with varying trophic coherence, 
 as described in the main text.
 (a) Incidence against $T$ (smaller $T$ means more coherent networks) in the complex contagion model for several values of the contagion
 parameter $\alpha$, as shown. 
 (b) Incidence against $\alpha$ in the complex contagion model for several values of $T$.
 (c) Incidence against $T$ in the Amari-Hopfield neural-network model for several values of the stochasticity
 parameter $\beta$.
 (d) Incidence against $\beta$ in the Amari-Hopfield neural-network model for several values of $T$.
All networks have $N=1000$, $B=50$, and $\langle k\rangle=5$.
Averages are over $1000$ runs.
 }
 \label{fig_traj}
 \end{figure*}

 \begin{figure*}
 \includegraphics[scale=1]{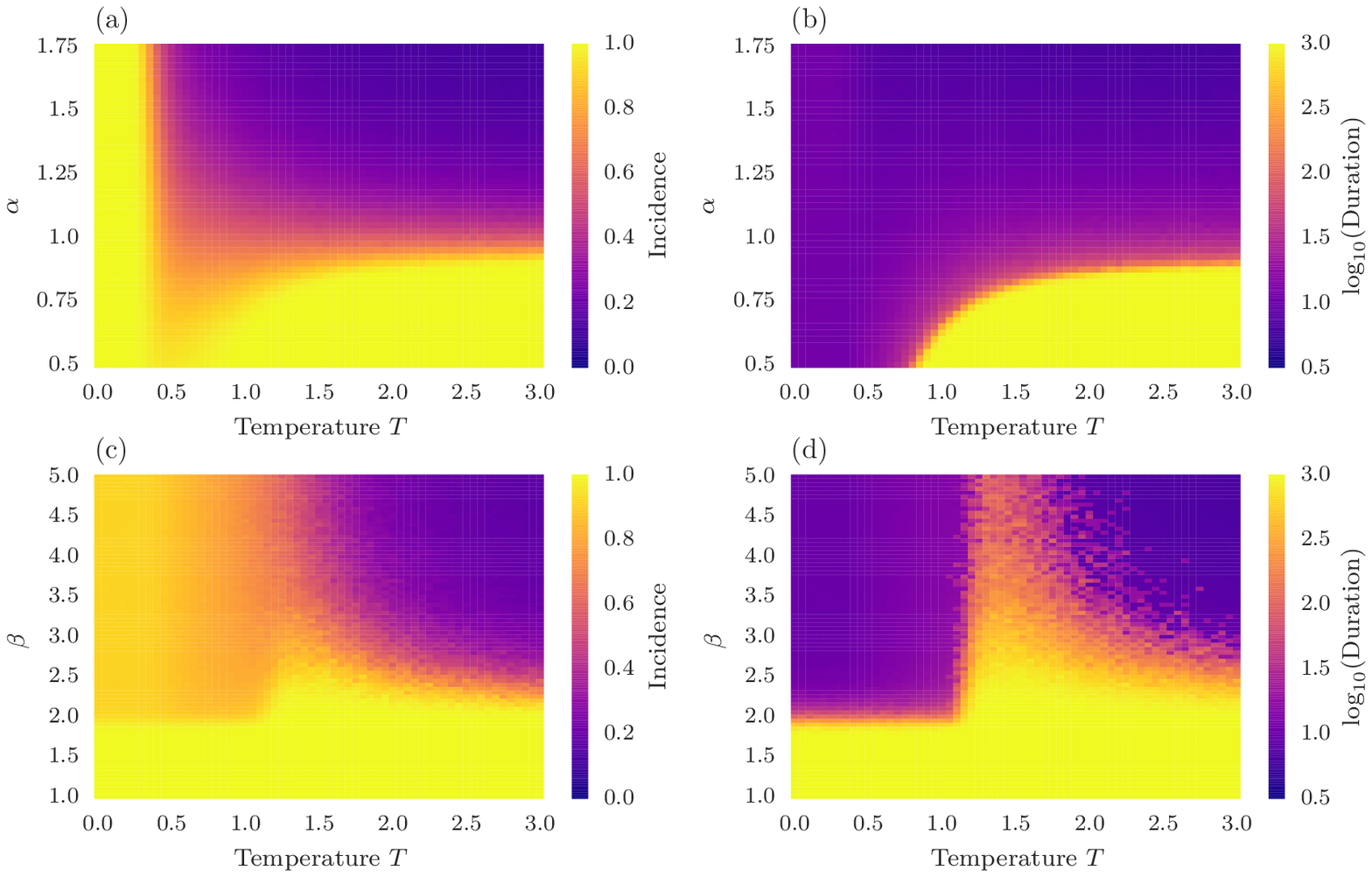}%
 \caption{
Heat-maps showing 
 average values of Incidence and of the common logarithm of Duration on a colour scale; 
 results are from Monte Carlo simulations of the two 
 spreading models on networks with varying trophic coherence, as set by $T$.
 (a) and (b) Complex contagion model, where $\alpha$ is the contagion parameter.
 (c) and (d) Amari-Hopfield neural-network model, where $\beta$ is the stochasticity parameter.
 All networks have $N=1000$, $B=50$, and $\langle k\rangle=5$.
 Averages are over $100$ runs.
 }
 \label{fig_heatmaps}
 \end{figure*}

\subsection{Spreading processes}
\noindent
The spreading of some form of activity through a system has been extensively studied in a wide variety of settings,
most notably in percolation theory.\cite{bollobas2006percolation,cohen2010complex}
Examples include epidemics,\cite{pastor-satorras_epidemic_2015} opinions,\cite{castellano2009nonlinear} 
forest-fires,\cite{drossel1992self} trophic cascades,\cite{power1992top} and avalanches of neural 
activity.\cite{beggs2007neuronal}
With a view to exploring the influence of trophic coherence on how activity of some kind spreads through a system, 
we consider two different paradigms: a model of complex contagion, and an Amari-Hopfield neural-network model.

\subsubsection{Complex contagion}
\noindent
Our first model 
is an adaptation of the standard Susceptible-Infected-Susceptible (SIS) epidemic model,\cite{pastor-satorras_epidemic_2015}
in which 
each node $i$ in a network is characterised at time $t$ by a binary variable $z_i(t)$, which
can be in either of two states: `susceptible' ($z_i(t)=S$) or `infected' ($z_i(t)=I$). Our version of this model will
take into account the phenomenon of ``complex contagion'', whereby the probability of an element becoming `infected' can be 
a non-linear function of the proportion of its neighbours who are already `infected'.
Centola has shown that social reinforcement plays a key role in the adoption of 
behaviour by participants in an online experiment,\cite{Centola1194} and theoretical research has recently highlighted how 
network clustering can influence this kind of spreading process.\cite{osullivan_mathematical_2015}
We consider that, at each time step $t$, each `susceptible' node $i$ has a probability of becoming `infected' given by
\begin{equation}
P[z_i(t+1)=I|z_i(t)=S]=f_i(t)^\alpha,
\label{eq_fa}
\end{equation}
where $f(t)$ is the fraction of $i$'s in-neighbours (i.e. of those nodes $j$ such that $a_{ij}=1$) which are in the `infected' 
state at time $t$ ($z_j(t)=I$), and $\alpha$ is a parameter which determines the kind of complex contagion.
A node which is `infected' at time $t$ automatically becomes `susceptible' again at time  $t+1$:
\begin{equation}
P[z_i(t+1)=S|z_i(t)=I]=1,
\end{equation}
and all nodes are updated in parallel.

The essence of complex contagion is captured also by the `q-voter model' of opinion dynamics,\cite{castellano2009nonlinear}
a generalization of the well-known voter model\cite{suchecki2005voter} in which the probability of an agent updating its state 
depends on there being a consensus among $q$ of its neighbours.

\subsubsection{Neural networks}
\noindent
The second model we consider is an Amari-Hopfield neural network, in which nodes are binary variables with states $v_i(t)=\pm 1$
representing the fact that, in a given time window, a neuron can either fire an action potential, 
or not.\cite{amari1972characteristics,hopfield1982neural,amit1992modeling}
Nodes are updated in 
parallel according to the probability
\begin{equation}
 P[v_i(t+1)=\pm1]=\frac{1}{2}\left\lbrace\pm\tanh\left[\beta h_i(t) \right]  +1\right\rbrace,
 \label{eq_Pv}
\end{equation}
where the field at $i$ is
\begin{equation}
 h_i(t)=\sum_j a_{ij}v_j(t),
\end{equation}
and the parameter $\beta$ sets the degree of stochasticity.
(Note that the expression for $h_i(t)$ usually takes into account 
the effects of ``synaptic weights'' which can be used to store information in the network, but for this work
we are considering all the weights to be equal.)

Although real neurons exhibit far richer behaviour than this simple model would suggest,
for many purposes collections of binary neurons are found to yield results qualitatively similar 
to those of more realistic models.\cite{abbott1990model}

\section{Results}
\noindent
In order to investigate numerically the effects of trophic coherence on spreading processes, we generate networks with given number 
of nodes $N$, 
basal nodes $B$, and edges $L$ as specified above, and carry out Monte Carlo runs of each of our dynamical models for different
values of the parameter $T$ -- i.e. for different degrees of trophic coherence.
For the complex contagion model, the initial condition is to set all nodes to `susceptible' except for the basal nodes, 
which are all `infected';
that is, $z_i(t=0)=S$ if $k^{in}>0$, and $z_i(t=0)=I$ if $k^{in}=0$.
For each run we measure the {\it Duration} of the infection, that is, the number of time steps until no nodes are `infected';
as well as the {\it Incidence}, or proportion of nodes which have at any 
time been in the `infected' state.

Figure \ref{fig_traj}(a) shows the mean Incidence against $T$ for various values of $\alpha$.
On highly coherent networks ($T\simeq 0$)
the infection spreads to the whole system for any $\alpha$. On less coherent topologies, however,
whether contagion is sub- or super-linear
has a strong influence on spreading: for $\alpha>1$ the infection only reaches a fraction of the network, while
for $\alpha<1$ the effect of coherence on Incidence is non-monotonic.
In Fig. \ref{fig_traj}(b), where the mean Incidence is plotted against $\alpha$ for different values of $T$, 
we can see how the effect of $\alpha$ on spreading is modulated by topology, becoming less severe the more 
coherent the networks.
Hence, it is the interplay of both the trophic coherence of the underlying network, 
and the form of the infection probability, which determines whether the infection can spread.

We also perform a similar investigation of the Amari-Hopfield neural model on networks generated in the same way. Now we set 
all the basal nodes initially to `firing' [i.e. $v_{i}(t=0)=1$ if $k_i^{in}=0$], and all other nodes to `not firing'
[$v_{i}(t=0)=-1$ if $k_i^{in}>0$]. The Duration is now the number of time steps to ensue before all the nodes 
are in the `not firing' state, and the Incidence is the proportion of nodes which at any moment during this period adopted the
`firing' state.
As with the infection, whether this pulse will propagate throughout the whole network
is determined by both the neural dynamics, as parametrised by $\beta$; and the trophic coherence of the network.
Figure \ref{fig_traj}(c) shows the mean Incidence against $T$ for several values of $\beta$, while Fig. \ref{fig_traj}(d)
has $\beta$ on the x-axis for various values of $T$.
Despite the different dynamics, the curves bear a resemblance to the cases in panels (a) and (b).
In both cases, a high trophic coherence ($T\simeq0$) can ensure that the pulse of activity will reach most of the network
irrespectively of other parameters, whereas if the network is incoherent ($T\gg0$) propagation requires low $\alpha$ (for the complex 
contagion model) or low $\beta$ (in the neural network).

 \begin{figure*}
 \includegraphics[scale=1]{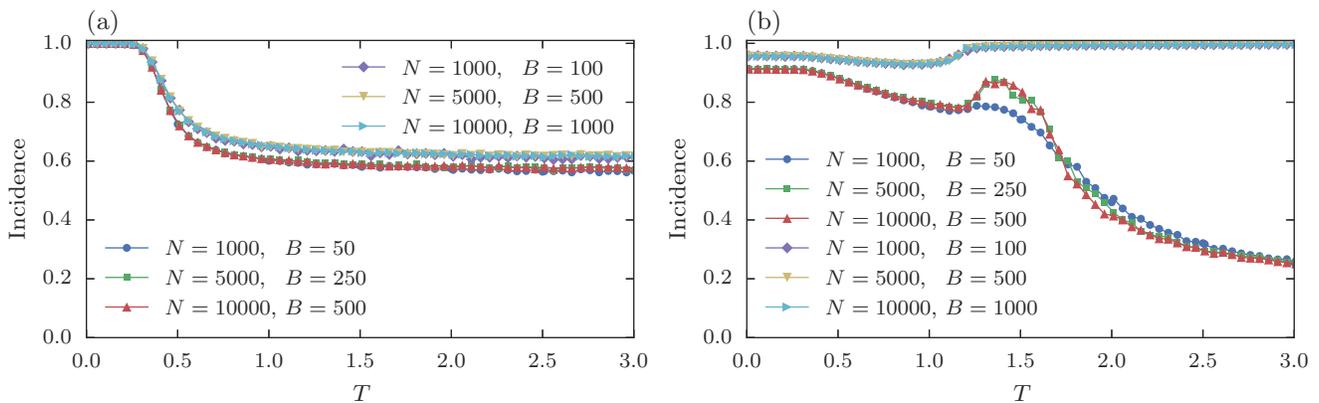}%
 \caption{Average Incidence values from Monte Carlo simulations of the two spreading models on networks with varying trophic coherence, 
 as described in the main text.
 (a) Incidence against $T$ (smaller $T$ means more coherent networks) in the complex contagion model for $\alpha=1$. 
 (b) Incidence against $T$ in the Amari-Hopfield neural-network model for $\beta=3$.
 Symbols indicate different network sizes ($N=1000$, $5000$ and $10000$) and proportions of basal nodes $B$ 
  ($N/B=10$ and $20$).
In all cases, the mean degree is $\langle k\rangle=5$.
Averages are over $1000$ runs.
}
 \label{fig_rev_traj}
 \end{figure*}

Figure \ref{fig_heatmaps} displays heat-maps for the complex contagion case [panels (a) and (b)] and the neural-network model 
[panels (c) and (d)]. Panels (a) and (c) show the mean Incidence against $T$ and the relevant model parameter ($\alpha$ 
for the complex contagion and $\beta$ for the neural network), while panels (b) and (d) show the logarithm of the Duration 
against the same parameters.
We run the simulations for a maximum of $10^3$ Monte Carlo steps, so Duration above this can mean either a long but eventually 
finite (transient) period of activation, or an endemic state in which a degree of activity remains indefinitely.

By comparing Incidence and Duration, we can discern that both models exhibit three qualitatively different regimes of behaviour:
at high $T$ and high $\alpha$ or $\beta$, activity dies out quickly without reaching most of the system;
at high $T$ and low $\alpha$ or $\beta$, activity spreads to the whole system and remains indefinitely;
at low $T$,
activity spreads to the whole system and then dies out quickly.
We can refer to these regimes as {\it inactive}, {\it endemic}, and {\it pulsing}, respectively.
The main qualitative difference between the behaviour of the two models regards the endemic regime.
In the complex contagion case, this regime is confined to sufficiently incoherent networks, and its range increases monotonically 
with $T$.
In the neural network it occurs for any $T$ if $\beta\lesssim2$, and the range is non-monotonic with $T$, peaking 
at intermediate values of $T$.

Why does trophic coherence affect spreading processes as described, and in such similar ways for 
both kinds of dynamics? 
Let us consider first the case of complex contagion on 
a perfectly coherent network (low $T$), like the one 
in Fig. \ref{fig_pic1}(a). If the basal nodes are all initially infected, then we have from Eq. (\ref{eq_fa}) that 
in the next time step
the probability of infection for nodes at level $s=2$ is $P=1$, for any $\alpha$, and thus the infection moves up a level.
By the same process, one time step later the infection moves to level $s=3$, and continues to spread in this 
way until it has reached the whole system -- at which point the infection dies out.
On an incoherent network (high $T$), like the one in Fig. \ref{fig_pic1}(b),   
as the pulse of activity moves up the trophic levels, the fraction of infected in-neighbours affecting 
a given node $i$, $f_i$, becomes lower with increasing $s_i$. 
Hence, if the network is insufficiently coherent, the pulse will die out as it progresses up the levels, and only 
reach a finite fraction of the nodes.
This explains why the pulsing regime occurs at low $T$. 
According to Eq. (\ref{eq_Pv}), the above considerations apply also to the neural-network model at low $T$ 
when $\beta$ is sufficiently high 
that the probability of a node being activated when all its in-neighbours are active is $P\simeq 1$.

For the complex contagion to become endemic, given that nodes recover immediately after infection, 
there must be some degree of feedback. In other words, the network must have cycles. 
As Johnson \& Jones have shown,\cite{Johnson_spectra} the expected number of cycles in a network is a function of its 
trophic coherence, and for $q$ below a particular value (which depends on other topological properties), networks are 
almost always acyclic. This accounts for an endemic phase which grows in range with $T$. However, the extent of node
re-infection will depend on both the density of cycles, and the probability that a node becomes infected by a given
proportion of infected in-neighbours, as determined by Eq. (\ref{eq_fa}); re-infection is therefore more likely at 
lower $\alpha$.
Again this argument can be extended to the neural-network model, with a caveat. In the complex contagion model, 
a node must have at least one infected in-neighbour to become infected, so the endemic regime requires cycles.
In the neural network, however, for any 
finite $\beta$ there exists a probability of spontaneous node activation. For low enough $\beta$, 
the system enters the standard paramagnetic (or memoryless) phase of the model, with continuous, random activation 
of nodes. This explains why the endemic regime of the neural model extends to the full range of $T$ for low $\beta$.

Finally, the non-monotonic dependence of the endemic regime with $T$ in the neural model seems to be caused by a 
balance between the two mechanisms we have described for activity propagation: a rapid pulse which can travel on coherent 
networks, and the reverberation allowed for by the cycles of incoherent ones.
It is perhaps noteworthy that this effect of feedback in the neural model is similar to the 
mechanism of `cluster reverberation' put forward to explain short-term memory.\cite{johnson2013robust}

To conclude we look into the effects of network size and of the proportion of basal nodes. Figure \ref{fig_rev_traj}
shows how Incidence depends on $T$ when we fix $\alpha=1$ for the complex contagion model [Fig. \ref{fig_rev_traj}(a)],
and $\beta=3$ in the neural network [Fig. \ref{fig_rev_traj}(b)]. Results are presented for three network sizes 
($N=1000$, $5000$ and $10000$), and two ratios $N/B=10$ and $20$.
In the complex contagion model the lines for different $N$ but fixed basal ratio collapse, and there is only a small effect of the 
basal ratio on the Incidence at high $T$. In the neural network there is a much more pronounced influence of the 
proportion of basal nodes: at high $T$, a ratio $N/B=10$ allows for spreading to reach the whole system when this is not 
possible if $N/B=20$. This may be a consequence of the dependence of the mean trophic level on this ratio,\cite{Johnson_spectra}
which for random graphs has an expected value $\langle s\rangle = N/B$.
When $N/B=20$, the non-monotonicity of Incidence with $T$ is also exacerbated slightly by $N$.

\section{Discussion}
 \noindent
We have shown that the trophic coherence of directed networks can have an important influence on spreading processes 
taking place thereon.
In particular,
our numerical investigation of two seemingly quite different dynamics -- one a model inspired by epidemics, the 
other a neural-network model originally put forward to explain associative memory -- 
indicates that this topological feature 
is relevant for
any system
in which some kind of signal is transmitted between elements in such as way that these signals interact.
We do not yet have an analytical theory able to describe spreading as a function of trophic coherence, 
but it is clear that such a theory should take into account two effects: the transmission of pulses of synchronous 
activity which can occur on highly coherent topologies; and the maintenance of endemic states 
enabled by feedback loops on incoherent networks. 
Trophic coherence has already been shown to play an important role in determining various features of 
directed networks, such as linear stability,\cite{Johnson_trophic} feedback,\cite{Johnson_spectra} and 
intervality.\cite{Vir_Chaos} We add here to such work by showing that spreading processes are also strongly 
influenced by this recently identified topological feature, and submit that more research is required to 
determine its relationship to other network properties, to build a generalised understanding of its 
bearing on dynamical processes, and to discover by what mechanisms non-trivial coherence or incoherence comes about 
in nature.

\begin{acknowledgments}
J. K. was supported by the EPSRC under grant EP/IO1358X/1. 
We are grateful to 
Miguel A. Mu{\~n}oz, Virginia Dom{\'i}nguez-Garc{\'i}a, and Nick S. Jones for feedback on the manuscript
as well as innumerable conversations
of great use and enjoyment.
\end{acknowledgments}






\end{document}